\documentclass[a4paper]{panl}
\usepackage{cite}
\usepackage{wrapfig}
\usepackage{graphicx}
\usepackage{amssymb}
\usepackage{amsfonts}
\usepackage{amsmath}
\usepackage{mathrsfs}
\usepackage{longtable}
\usepackage{rotating}
\usepackage{lscape}
\usepackage{epsfig}
\usepackage{multirow}

\originalTeX
\begin{document}
\issuearea{Physics of Elementary Particles and Atomic Nuclei. Theory} 

\title{Hyperons from Bi+Bi collisions at MPD-NICA: Preliminary analysis of production at generation, simulation and reconstruction level}
\maketitle
\authors{
Alejandro Ayala$^{a,b}$\, 
Eleazar Cuautle$^{a}$\,
Isabel Dom\'{\i}nguez$^{c}$\, 
M. Rodr\'{\i}guez-Cahuantzi$^{d}$ }
\authors{
Ivonne Maldonado$^{c}$\footnote{ivonne.alicia.maldonado@gmail.com}\,
Mar\'{\i}a Elena Tejeda-Yeomans$^{e}$}
\from{$^{a}$\, Instituto de Ciencias Nucleares, Universidad Nacional Aut\'onoma de M\'exico, Apartado Postal 70-543, C.P. 04510, CDMX, M\'exico}
\vspace{-5mm}
\from{$^{b}$\, Centre  for  Theoretical  and  Mathematical  Physics,  and  Department  of  Physics,University  of  Cape  Town,  Rondebosch  7700,  South  Africa}
\vspace{-5mm}
\from{$^{c}$\, Facultad de Ciencias F\'{\i}sico-Matem\'aticas, Universidad Aut\'onoma de Sinaloa, Avenida de las Am\'ericas y Boulevard Universitarios, Ciudad Universitaria, C.P. 80000, Culiac\'an, Sinaloa, M\'exico }
\vspace{-5mm}
\from{$^{d}$\, Facultad  de  Ciencias  F\'{\i}sico  Matem\'aticas,  Benem\'erita  Universidad  Aut\'onoma  de  Puebla, Av.   San  Claudio  y  18  Sur,  Edif.   EMA3-231,  Ciudad  Universitaria  72570,  Puebla,  Mexico.}
\vspace{-5mm}
\from{$^{e}$\, Facultad de Ciencias-CUICBAS, Universidad de Colima, Bernal  D\'{\i}az  del  Castillo  No.   340,  Col.   Villas  San  Sebasti\'an,  28045  Colima,  Mexico.}


\begin{abstract}


An important observable to understand the properties of the matter produced in heavy-ion collisions is its strangeness content. Recent experimental results show that in semi-central collisions, the $\Lambda$ and $\bar{\Lambda}$ global polarization show differences that increase at low energies. This behaviour has been described using a model where these particles may be produced from two distinct density zones in the collision region: the \textit{core} and the \textit{corona} where QGP processes and p + p like reactions, respectively, are mainly at work. Using this idea, the polarization can be influenced by the relative abundance of these particles coming from either regions. In this work we show how to test this model in the MPD experiment. 

\end{abstract}
\vspace*{6pt}

PACS: 14.20.$+$Jn 

\section{Introduction}\label{sec:intro}

The system created in heavy-ion collisions provides an environment where we can study the spin polarization properties of the produced hadrons.
In recent years, this observable has drawn a great deal of attention due to the possibility that the spin of hadrons may be aligned along the global vorticity produced in non-central collisions~\cite{Becattini:2007sr,Becattini:2016gvu}. Among these hadrons, $\Lambda$ and $\bar{\Lambda}$ play an important role due to their self-analyzing polarization properties. 

ALICE~\cite{Acharya:2019ryw} and the STAR Beam Energy Scan (BES)~ \cite{STAR:2017ckg,Adam:2018ivw} have measured the $\Lambda$ and $\bar{\Lambda}$ global polarization as a function of the collision energy. In particular, the STAR BES found that both polarizations increase
as the collision energy decreases, with the $\bar{\Lambda}$ polarization becoming larger than the $\Lambda$ polarization. Recently, this behaviour has been studied assuming that in non-central heavy-ion collisions, these hyperons can be produced both from a high-density core and from a less dense corona~\cite{Ayala:2020soy,Ayala:2020vyi,Ayala:2001jp}. 
The $\bar{\Lambda}$ global polarization is amplified despite that the intrinsic $\Lambda$ polarization is larger than the intrinsic $\bar{\Lambda}$ polarization~\cite{Ayala:2020ndx,Ayala:2019iin}, when a larger abundance of $\Lambda$s with respect to $\bar{\Lambda}$s in the corona is combined together with a smaller number of $\Lambda$s from the core compared to those coming from the corona.
In this work we present a technique to test this idea within the Multi-Purpose Detector (MPD) at the Nuclotron-based Ion Collider fAcility (NICA) located at the Joint Institute for Nuclear Research (JINR)~\cite{Abraamyan:2011zz,Golovatyuk:2016zps} designed to study heavy-ion collisions in the center of mass energy range $\sqrt{s_{NN}}=\{4, 11\}$ GeV. In order to test the model, 
we measure the hyperon global polarization for different centrality sets of data and compare them with the assumptions made in the model. The measurement requires analysis at three different steps: MC event generation, the transport through detector and the reconstruction within the MPDroot framework, for both the hyperon identification and event plane determination.

This work is organized as follows: In Sec.~2 we review how the standard global polarization measurement is performed. In Sec.~3, we describe the analyzed data and in Sec.~4  we describe the preliminary results of hyperon reconstruction and angular distributions required to obtain the global polarization. We finally conclude and summarize in Sec.~5.


\label{sec:HyperonGlobalPolarization}
\section{Hyperon Global Polarization}

\begin{figure}[!h]
\begin{center}
\includegraphics[width=60mm]{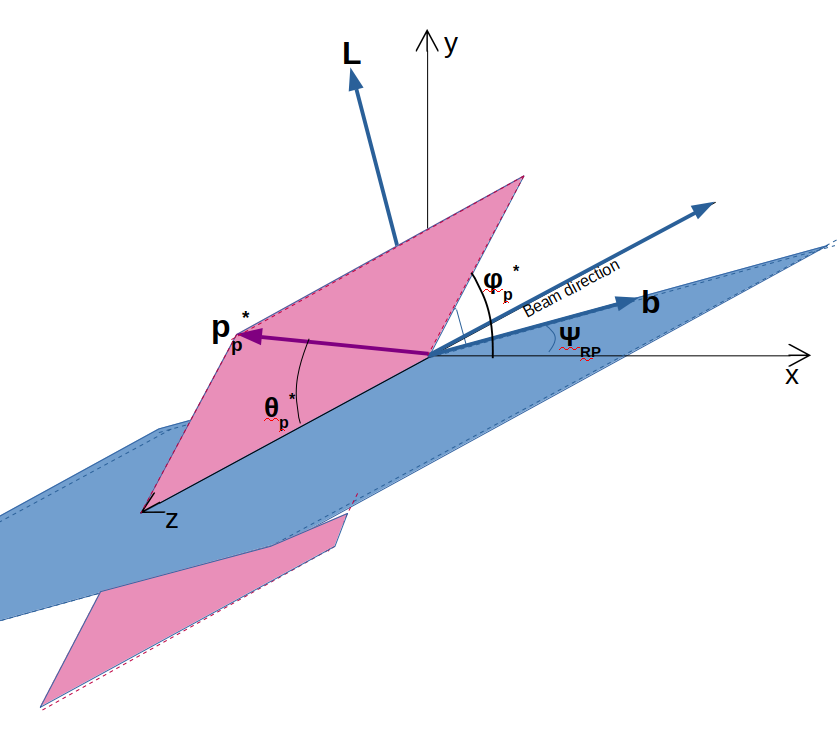}\hspace{10mm} 
	\begin{minipage}[b]{50mm}
		\caption{Diagram that relates the laboratory frame and the hyperon rest frame. The reaction plane is defined by the impact parameter $\hat{\textbf{b}}$ and the beam direction $\hat{\textbf{p}}_{beam}$.}
\end{minipage}
\end{center}
\labelf{figura0}
\end{figure}

The hyperon global polarization is measured relative to the system's orbital angular momentum $\hat{\textbf{L}}$, which is perpendicular to the reaction plane and defined by  $\hat{\textbf{L}} = \hat{\textbf{b}} \times \hat{\textbf{p}}_{beam}$, namely, the cross product of the impact parameter $\hat{\textbf{b}}$ and the beam direction $\hat{\textbf{p}}_{beam}$. The angular distribution of the hyperon decay products relative to $\hat{\textbf{L}}$ is given by
\begin{eqnarray}
\frac{dN}{d \Omega^{*}} = \frac{N}{4\pi}(1 + \alpha_H \mathscr{P}_H\cos{\theta^{*}}),
\label{dist}
\end{eqnarray}
where $N$ is the number of particles, $\mathscr{P}_{H}$ is the hyperon global polarization, $\alpha_H$ is the hyperon decay parameter ($\alpha_{\Lambda} = 0.642 \pm 0.013$)~\cite{PhysRevD.98.030001} and $\theta^{*}$ is the angle in the hyperon rest frame between the system's orbital angular momentum $\hat{\textbf{L}}$ and the three-momentum of the baryon produced by the hyperon decay $\mathbf{p}^*_p$. 

Polarization in equation~(\ref{dist}) can be rewritten in terms of the reaction plane angle $\Psi_{RP}$ and the azimuthal angle $\phi^*_p$ of the hyperon decay baryon three-momentum in the hyperon rest frame, by means of the trigonometric relation between the angles in the laboratory frame and the $\Lambda$ rest frame given by $\cos{\theta^*} = \sin{\theta^*_p}\sin{(\phi^*_p  - \Psi_{RP})}$, where $\theta^*_p$ refers to the angle between the three-momentum of the baryon produced by the hyperon decay and the $z$ laboratory frame axis as depicted in Fig.~ \ref{figura0}. This results in an expression for the hyperon global polarization given by  \begin{eqnarray}
\mathscr{P}_{H} = \frac{8}{\pi \alpha_H}\langle\sin{(\phi^*_p - \Psi_{RP})} \rangle.
\label{pol}
\end{eqnarray}

\label{sec:Data analyzed}
\section{Data analysis}

For this study we generate 100,000 Bi + Bi events at $\sqrt{s_{NN}} = 11$ GeV for different centrality sets of data:
\begin{itemize}
    \item Minimum Bias.
    \item Central collisions, with $b < 4$ fm.
    \item Semi-central collisions, with $b \in (6,8)$ fm.
    \item Peripheral collisions, with $b> 10$ fm.
    \end{itemize}
We use UrQMD and Geant3 for generation and transport through the detector. The transport and reconstruction usually considers the TPC, TOF, EMC and ZDC detectors, however for this part of the analysis we only consider the TPC, that is, the main tracking detector in the MPD barrel which covers the mid-rapidity region $|\eta|<1.2$ and $p_{\textit{T}} > 100$ MeV/c ~\cite{Averyanov:2017oec}. 

We reconstruct the hyperons through its weak decay topology into proton (antiproton) and corresponding charged pion using the TPC. We analyze $\Lambda$ and $\bar{\Lambda}$ at three steps: 
generation, simulation and reconstruction. For this purpose we define the particles analyzed at each level.
\begin{itemize}
    \item Monte Carlo data (MC), $\Lambda$s and $\bar{\Lambda}$s produced with UrQMD. In addition we consider $\Lambda$s and $\bar{\Lambda}$s coming from decays of particles such as $\Omega$, $\Xi$ and $\Sigma$ to account for secondary interactions produced by GEANT3 with the different elements of the detector.
    \item Simulated data (sim), $\Lambda$ and $\bar{\Lambda}$ that can be identified by Monte Carlo association of the products of its charged decay and with transverse momentum $p_{\textit{T}}>0.001$ GeV$/c$ and $|\eta| < 1.3$, to be in the acceptance of the detector.
    \item Reconstructed data (rec) namely $\Lambda$ and $\bar{\Lambda}$ identified by the combination of secondary tracks of opposite identified charge, namely, p$^+$(p$^-$) and $\pi^-$($\pi^+$), together with background subtraction.
 \end{itemize}
Table~\ref{tab:table1} shows the number of $\Lambda$s and $\bar{\Lambda}$s per event for each different level of the analysis for the different sets of data. The abundance for each impact parameter range is also shown in Fig.~\ref{figura2}.
\begin{table}[!ht]
    \centering
    \begin{tabular}{|c|c|c|c|c|c|c|}
    \hline
    Data  & \multicolumn{2}{|l|}{Generated}  & \multicolumn{2}{|l|}{Simulated} & \multicolumn{2}{|l|}{Reconstructed} \\
    \hline
    \hline
    Sample & $\Lambda$ & $\bar{\Lambda}$& $\Lambda$ & $\bar{\Lambda}$ & $\Lambda$ & $\bar{\Lambda}$ \\
         \hline
    MB            & 11.8 & 0.22 & 6.36 & 0.14 & 0.66 & 0.02 \\
    $b < 4$ fm     & 50.6 & 0.74 & 28.2 & 0.47 & 3.78 & 0.06 \\
    $6 < b < 8$ fm & 24.0 & 0.45 & 13.1 & 0.28 & 1.16 & 0.04 \\
    $b > 10$ fm   & 2.12 & 0.07 & 1.10 & 0.04 & 0.05 & 0.004 \\
         \hline
    \end{tabular}
    \caption{Content of $\Lambda$ and $\bar{\Lambda}$ for the different datasets.}
    \label{tab:table1}
\end{table}
\begin{figure}[t]
\begin{center}
\includegraphics[width=127mm]{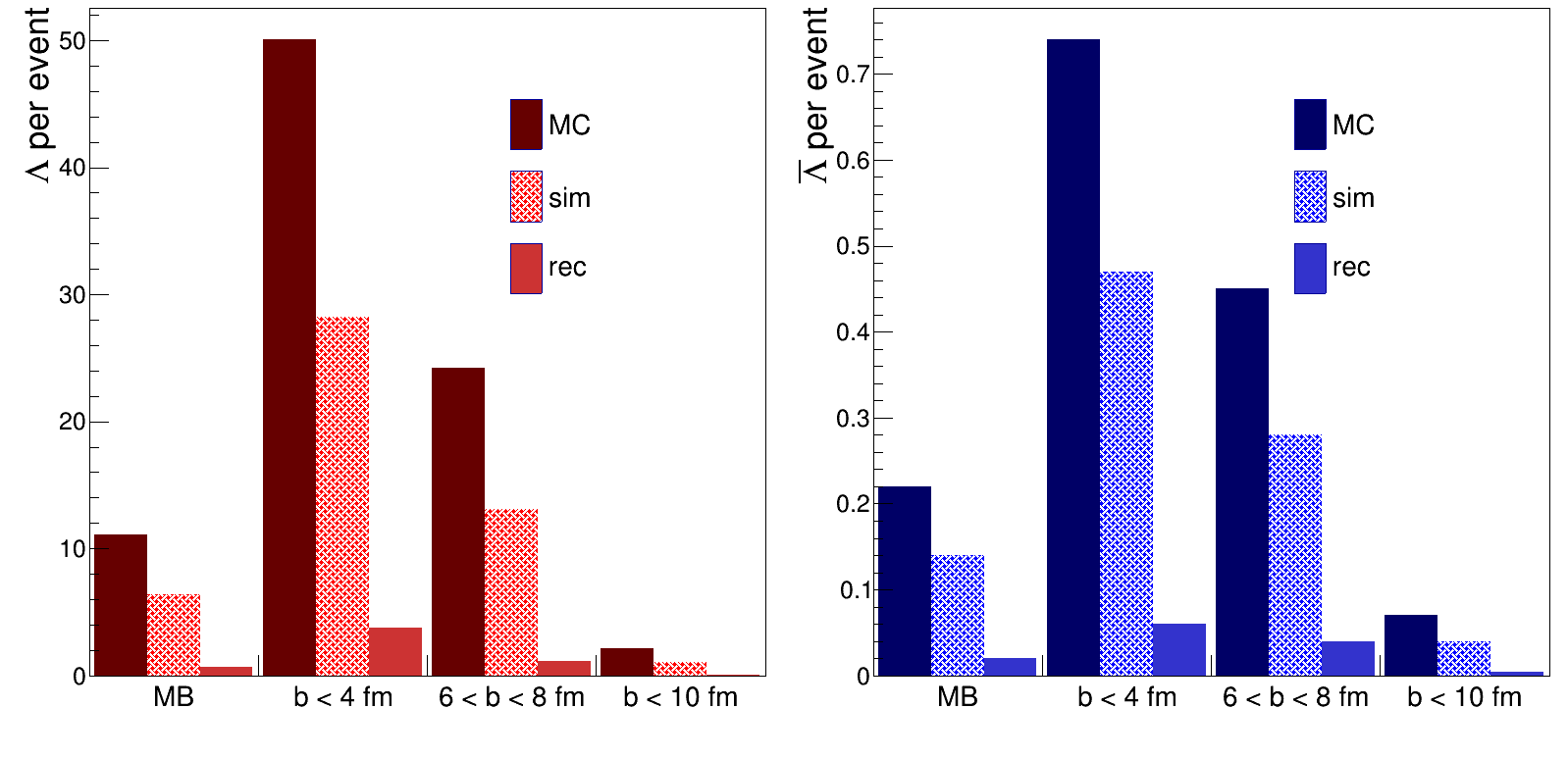} 
\vspace{-3mm}
\caption{Number of $\Lambda$s (left) and $\bar{\Lambda}$s (right) per event in each dataset.}
\end{center}
\labelf{figura2}
\vspace{-5mm}
\end{figure}


In addition we need to estimate the maximum efficiency $\varepsilon$ to be obtained in the reconstruction of particles so as to correct the effects of detector acceptance and resolution. The maximum efficiency is made identifying the reconstructed hyperons with the MC association. 
Figure~\ref{figura3} show the maximum efficiency for $\Lambda$ and $\bar{\Lambda}$ as a function of $p_{\textit{T}}$. We choose this variable in order to compare with the results of any other possible analysis. We observe that the efficiency is similar for the different sets of analyzed data. The maximum value is $\varepsilon\approx 0.3$ for $p_{\textit{T}}\approx 2$ GeV. To reproduce this efficiency, we require to improve the particle identification.  We should get this efficiency as a function of the variables to be analyzed, such as the azimuthal angle, to correct for detector effects.

\begin{figure}[t]
\begin{center}
\includegraphics[width=63.5mm]{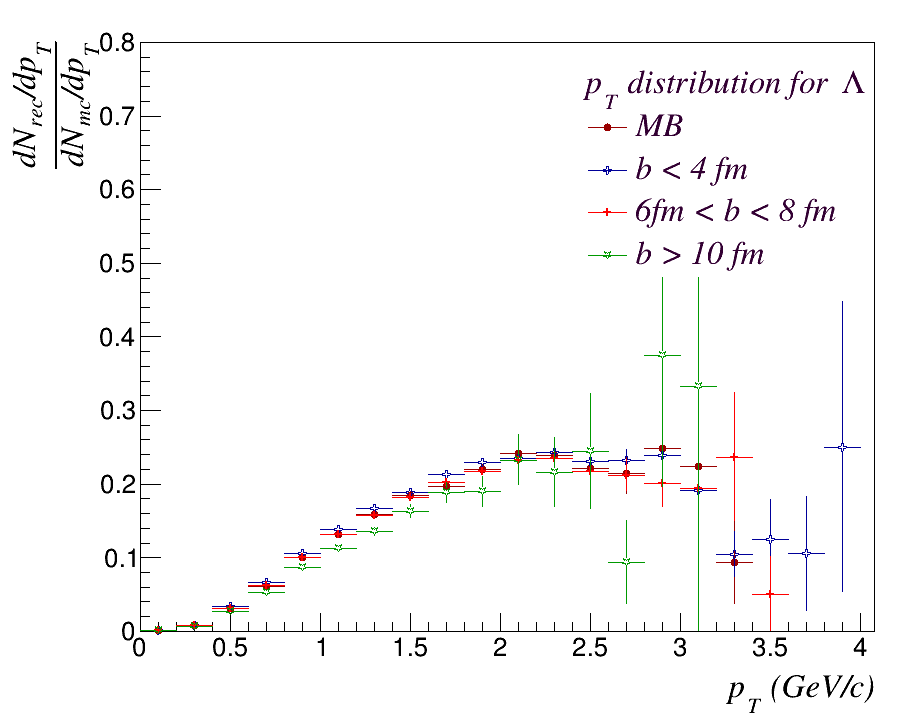} 
\includegraphics[width=63.5mm]{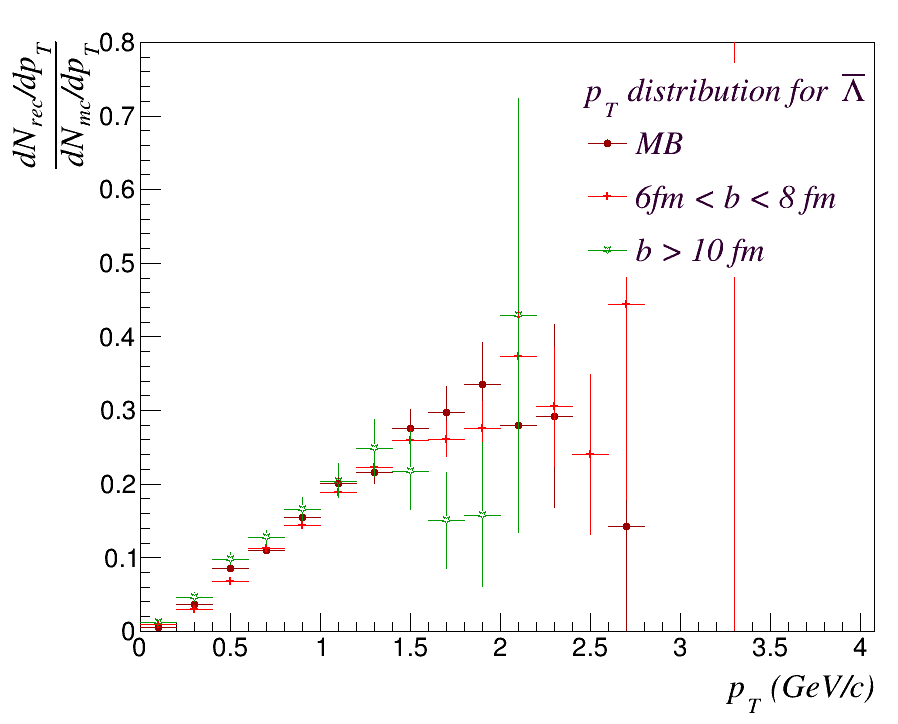}
\vspace{-3mm}
\caption{Reconstruction efficiency for $\Lambda$ (left) and $\bar{\Lambda}$ (right) in each data set. Both shows a rise in efficiency for $0 < p_{\textit{T}} < 2$ GeV/c.}
\end{center}
\labelf{figura3}
\vspace{-5mm}
\end{figure}

\label{sec:Reconstruction}
\section{Hyperon reconstruction}

We select $\Lambda$ and $\bar{\Lambda}$ using their weak decay topologies, as is shown in Fig.~\ref{figura4}. The $V^0$ finding procedure starts with the combination of each secondary track with every secondary track of opposite charge. To get the invariant mass, the selection of candidates is done using cuts on variables such as the produced baryon distance of closest approach to the primary vertex (DCA $p$-track and DCA $\pi$-track), the distance of closest approach between the two produced particle tracks (DCA V0) and the cosine of the angle between the $V^0$ reconstructed momentum and a vector $\textbf{R}$ joining the primary and secondary vertices (Cosine($\theta$)). The distribution of these variables as a function of invariant mass are shown in Fig.~\ref{figura5}. To distinguish between $\Lambda$ and $\bar{\Lambda}$ we use the asymmetry of the longitudinal momentum of product tracks in the rest frame of the hyperon given by the Armenteros-Podolanski variables\cite{doi:10.1080/14786440108520416}.
\begin{figure}[h!]
\begin{center}
\includegraphics[width=127mm]{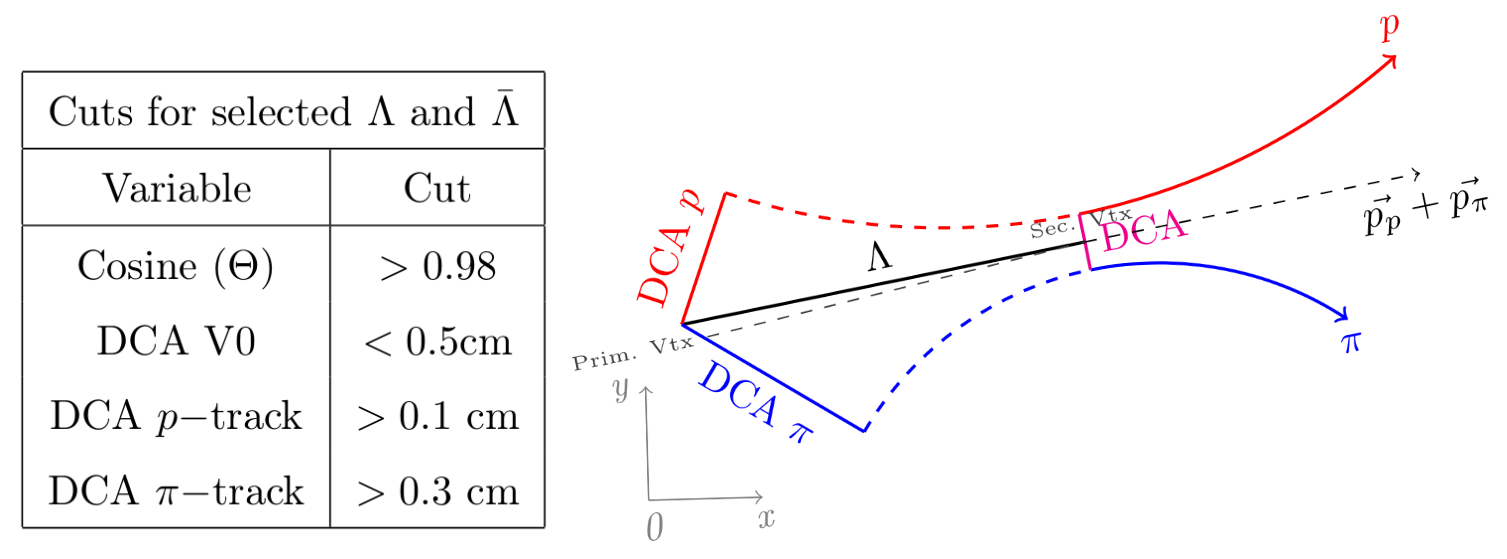} 
\vspace{-3mm}\caption{Topological reconstruction variables.}
\end{center}
\labelf{figura4}
\vspace{-5mm}
\end{figure}

\begin{figure}[h!]
\begin{center}
\includegraphics[width=63.5mm]{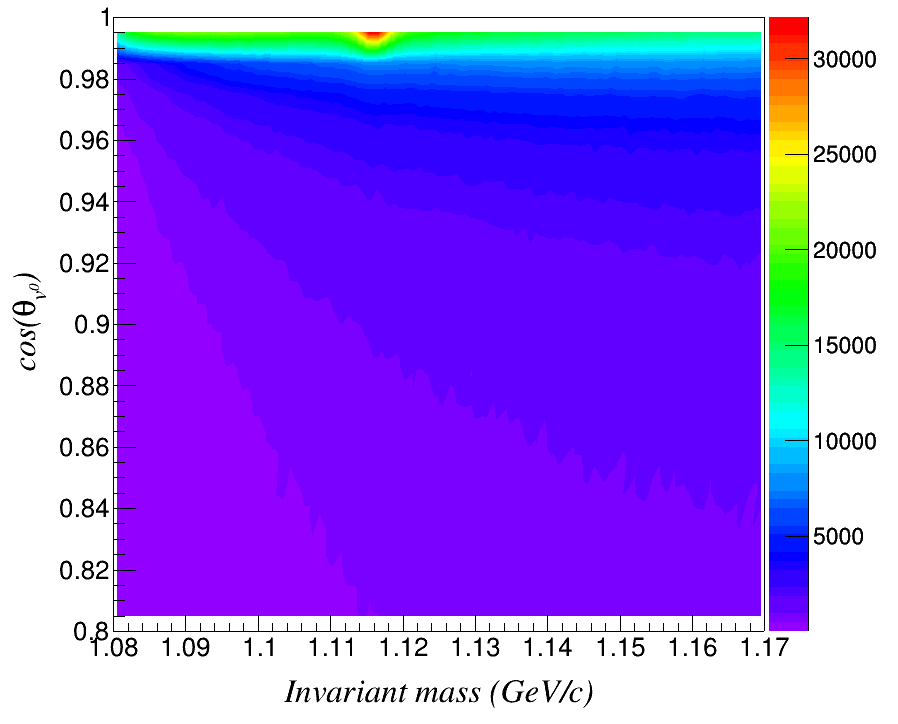} 
\includegraphics[width=63.5mm]{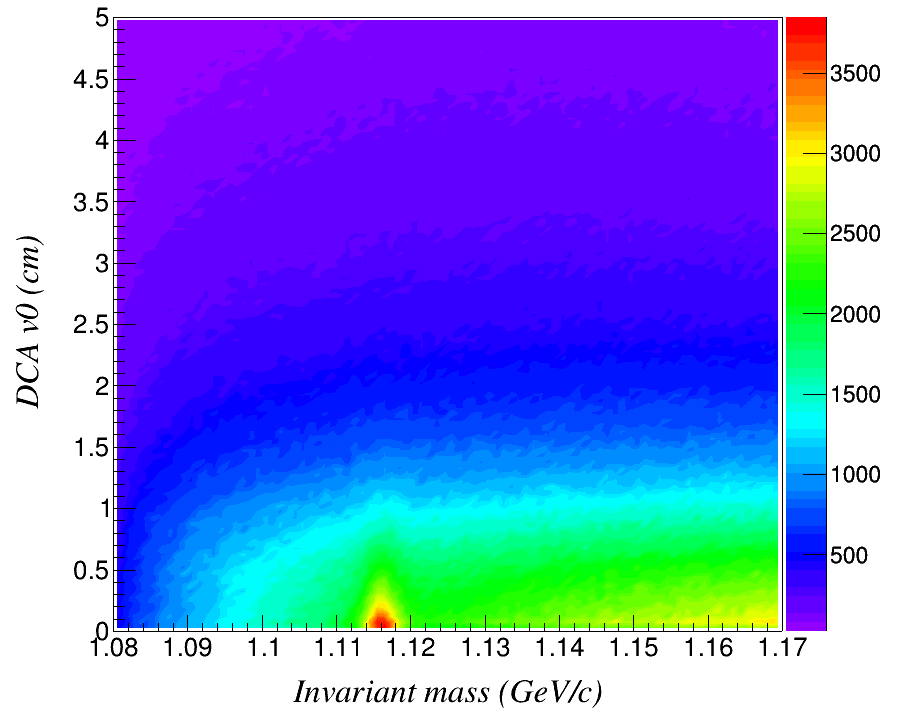}\\
\includegraphics[width=63.5mm]{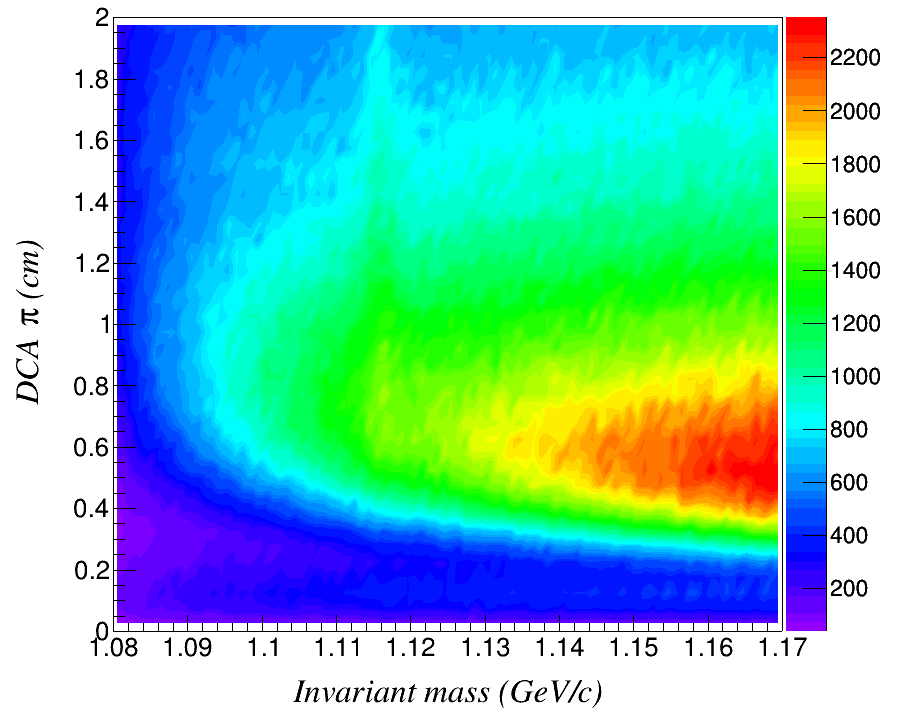}
\includegraphics[width=63.5mm]{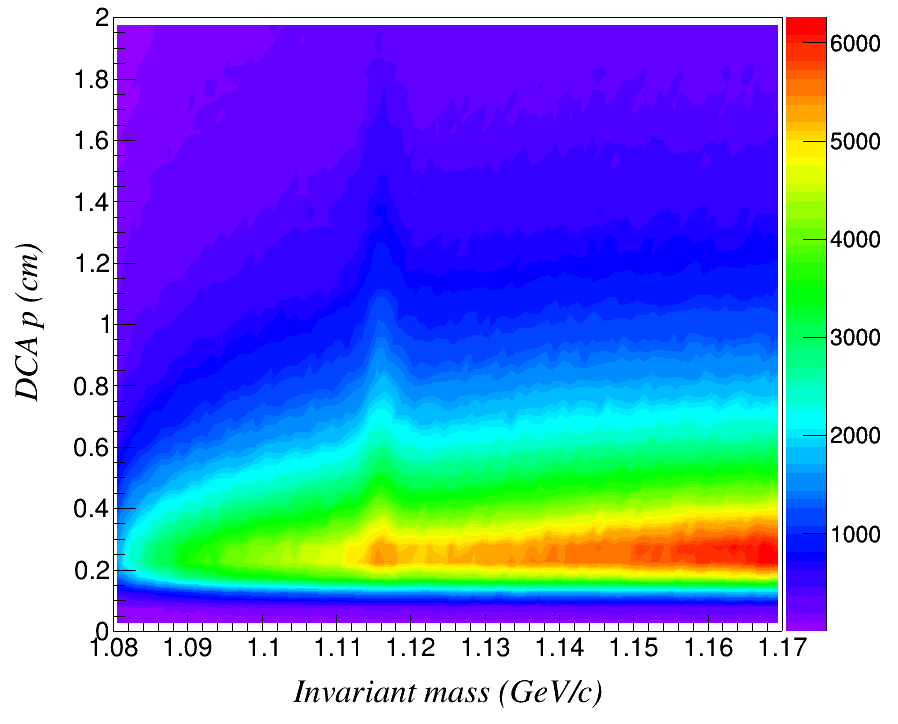}
\vspace{-3mm}\caption{The topological variables vs. the hyperon invariant mass.}
\end{center}
\labelf{figura5}
\vspace{-5mm}
\end{figure}
The invariant mass distributions for $\Lambda$ and $\bar{\Lambda}$ are obtained implementing the cuts shown in Fig.~\ref{figura5} for the different data sets shown in Fig.~\ref{figura6}. 
\begin{figure}[h!]
\begin{center}
\includegraphics[width=42.33mm]{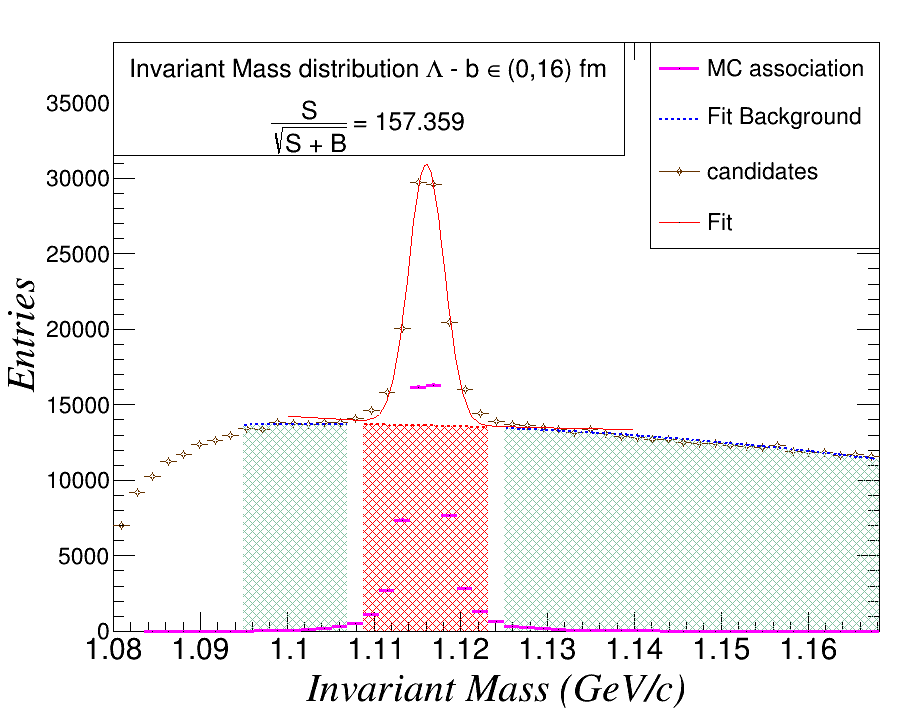} 
\includegraphics[width=42.33mm]{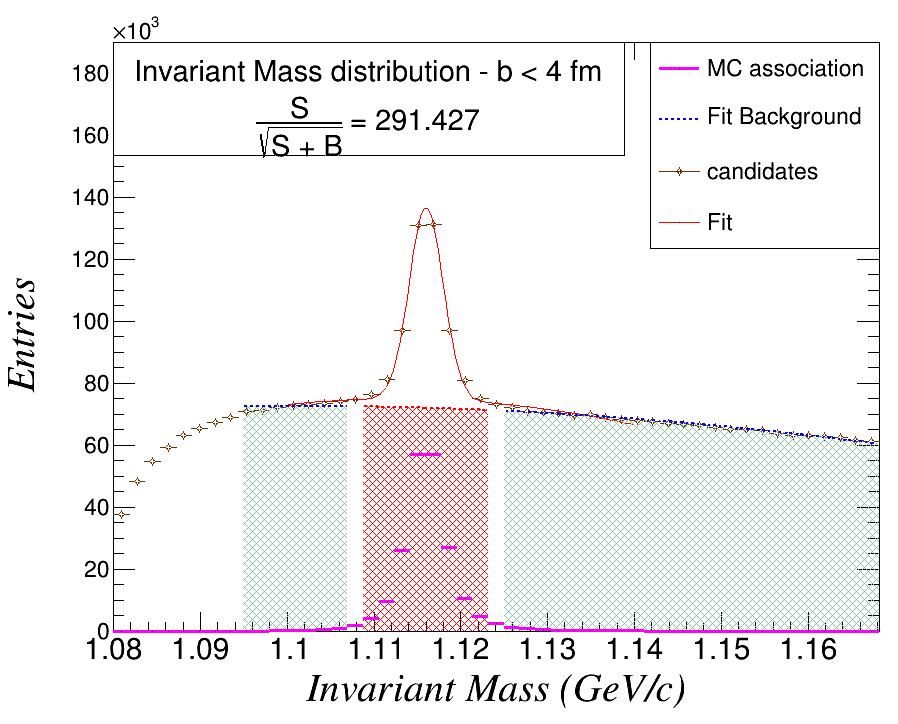}
\includegraphics[width=42.33mm]{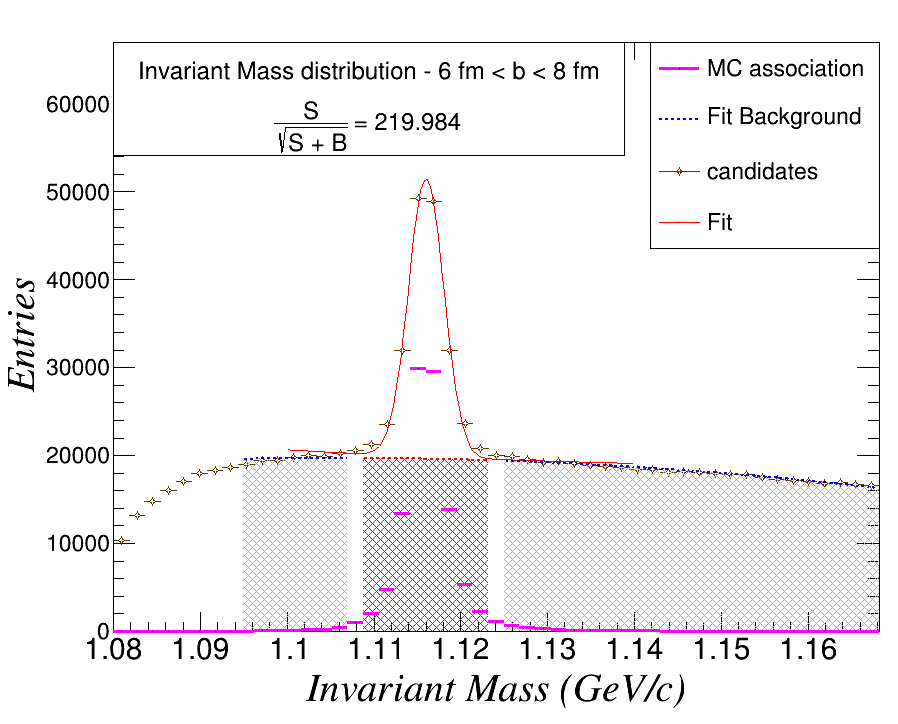}\\
\includegraphics[width=42.33mm]{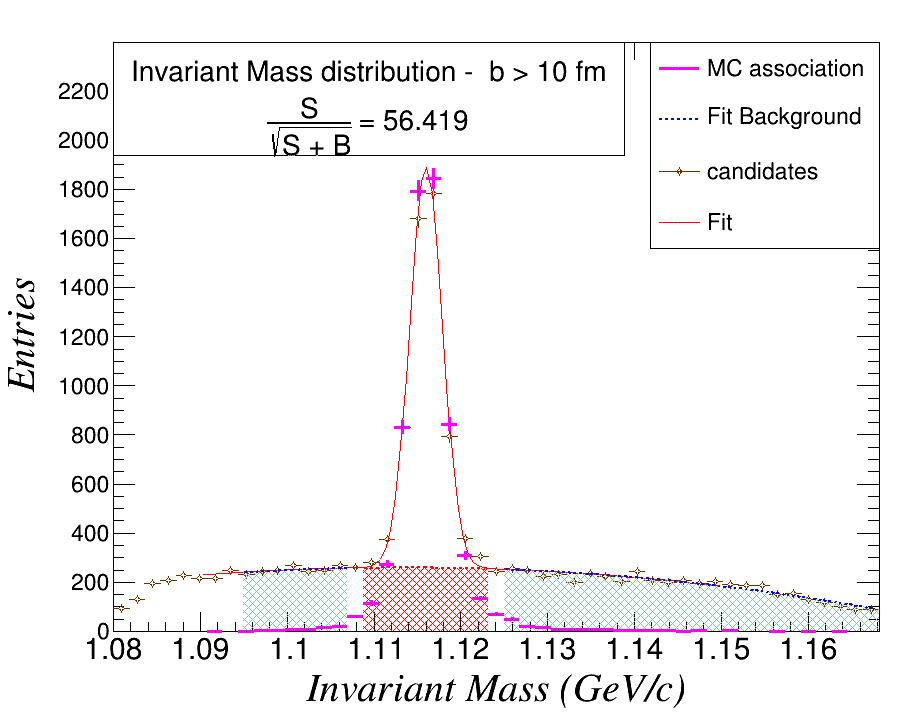}
\includegraphics[width=42.33mm]{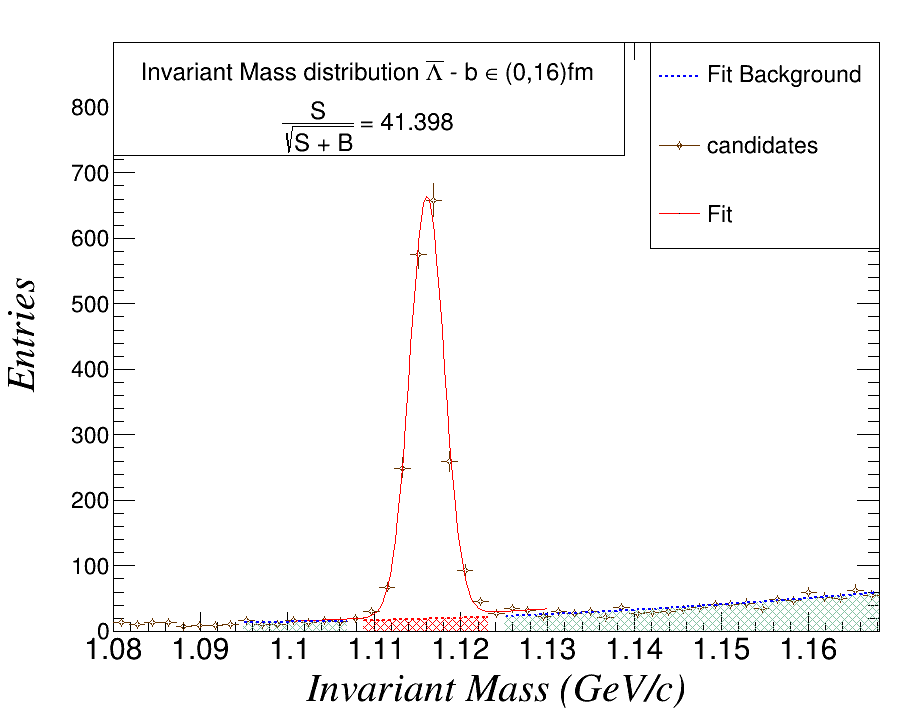} 
\includegraphics[width=42.33mm]{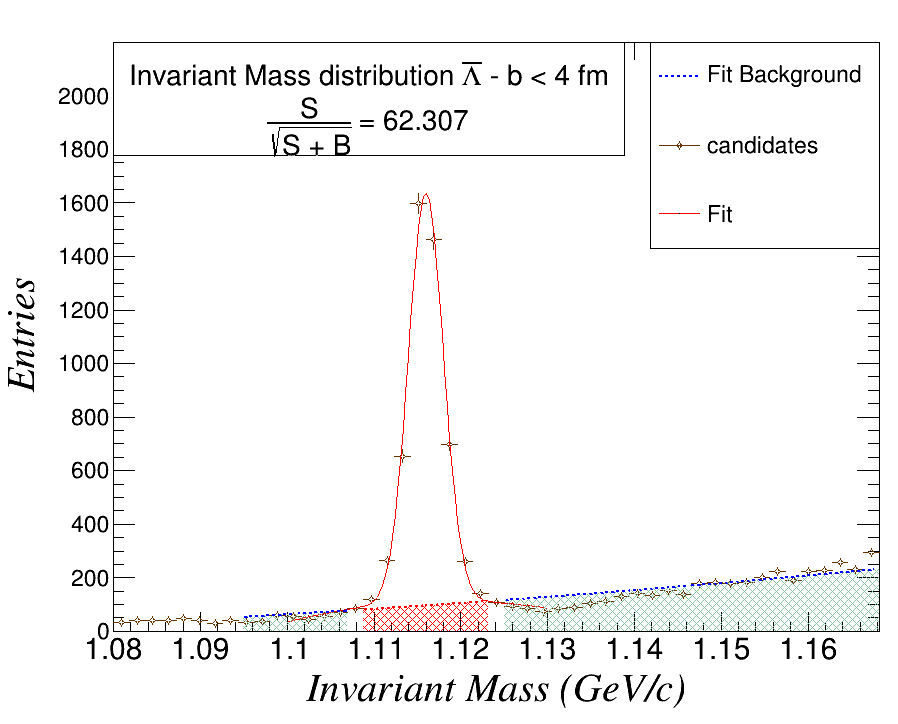}\\
\includegraphics[width=42.33mm]{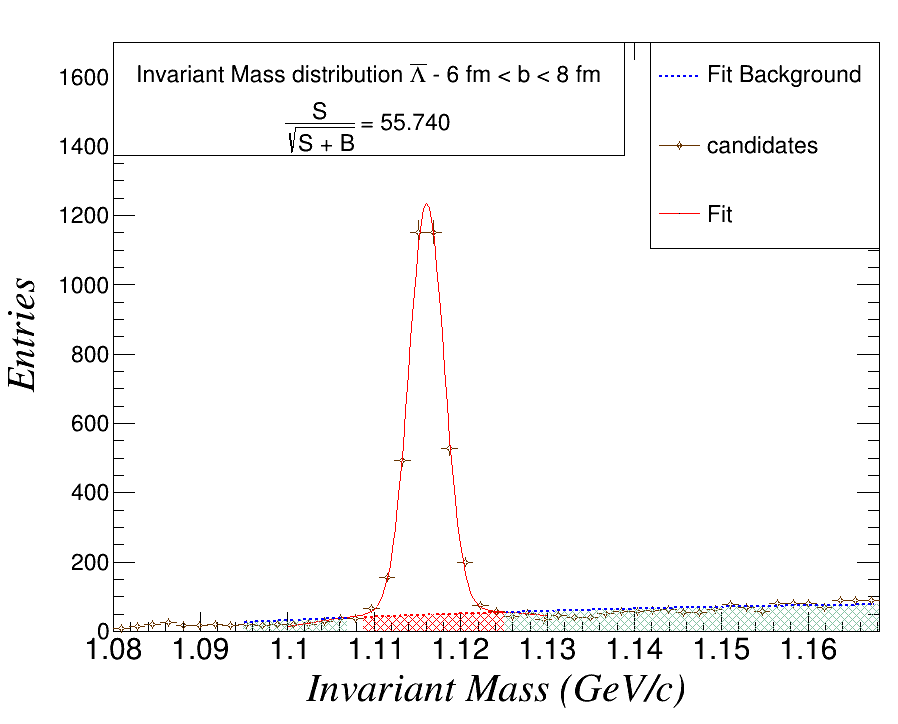}
\includegraphics[width=42.33mm]{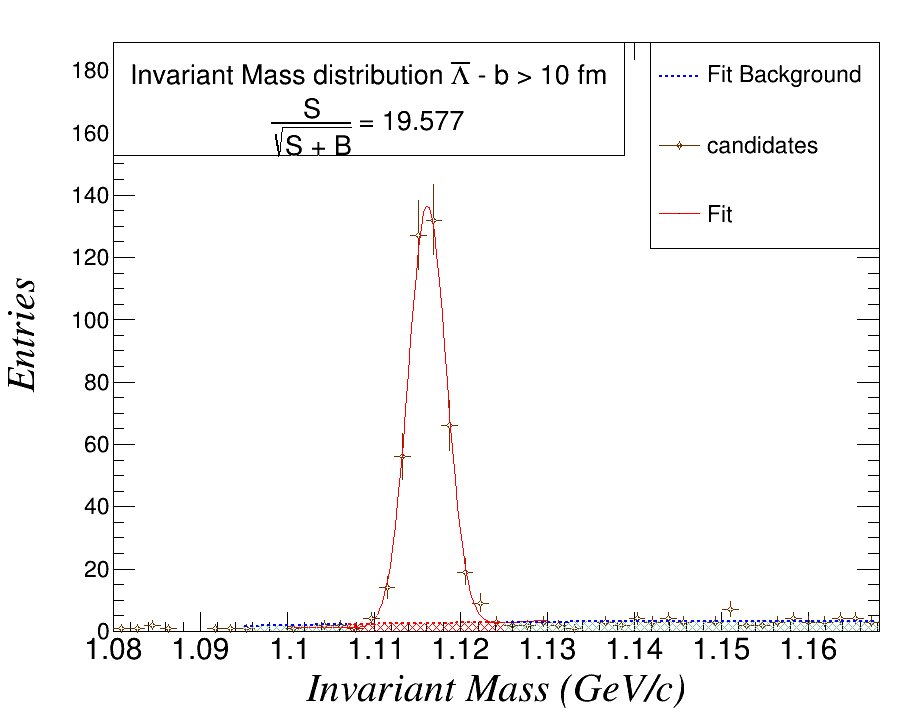}
\vspace{-3mm}\caption{Invariant mass distribution for $\Lambda$ and $\bar{\Lambda}$ obtained with the cuts in Fig.~\ref{figura4}. To improve the selection, we need to change these cuts maximizing the significance and using PID for decay particles.}
\end{center}
\labelf{figura6}
\vspace{-5mm}
\end{figure}
\begin{figure}[bh!]
\begin{center}
\includegraphics[width=55.5mm]{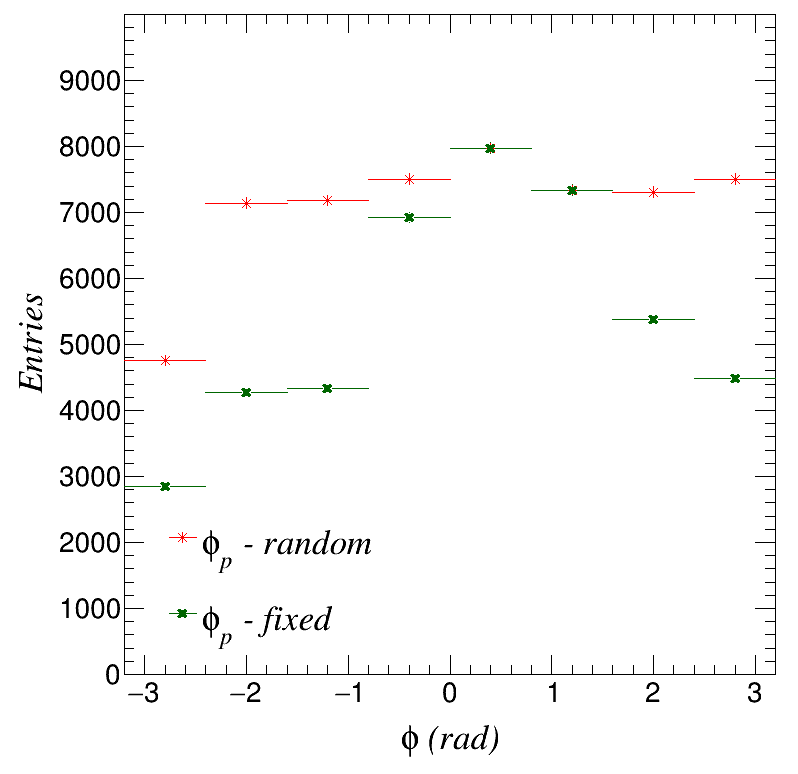} 
\includegraphics[width=55.5mm]{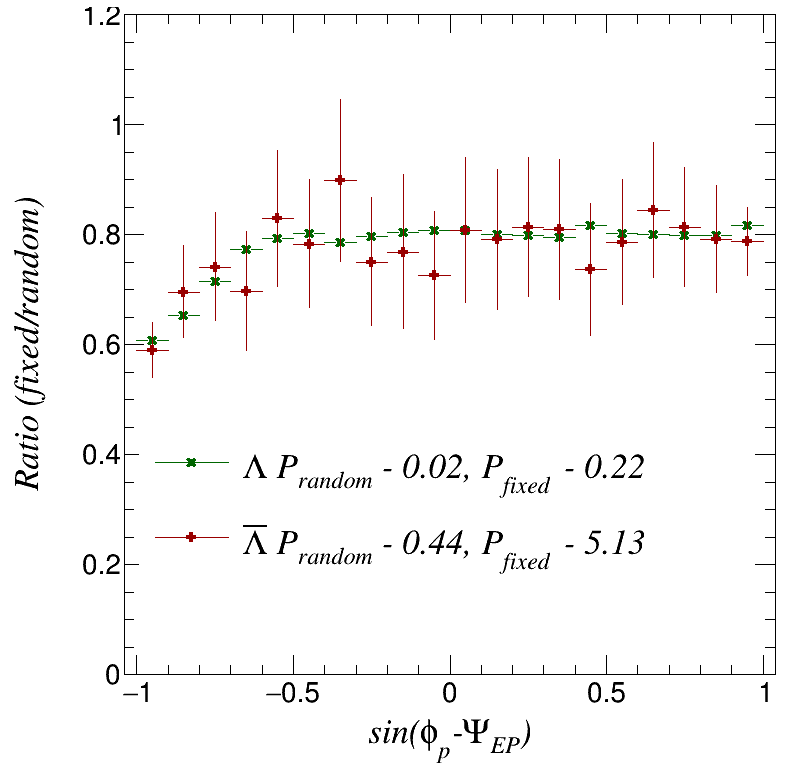}
\vspace{-3mm}\caption{(Left) Azimuthal baryon decay product distribution showing the homegeneous and inhomogeneus distributions. (Right) Estimation of the polarization showing that the distributions in the left panel lead to different estimations.}
\end{center}
\labelf{figura7}
\vspace{-5mm}
\end{figure}
The number of $\Lambda$s ($\bar{\Lambda}$s) is larger for central collisions and decreases for peripheral collisions, as is expected from the generated data.
Once the hyperons are identified, we compare the azimuthal angle of the baryon products with the event plane angle $\Psi_{RP}$. The event plane orientation is estimated in terms of the first order event plane angle $\Psi^{(1)}_{EP}$ and its resolution $R^{(1)}_{EP}$ using Eq.~(\ref{pol})~\cite{Abelev:2007zk}.

As a method to test our protocol to measure the hyperon polarization, we first obtain a set of hyperons for which we choose an inhomogeneous distribution of the azimuthal angle of the baryon decays so that we can compare with the MB set of unpolarized data and check for the differences. Also, for this part of the analysis, we use the reaction plane angle $\Psi_{RP}$, assigned by the generator during the transport through the detector. As a next step we plan to use the measured angle $\Psi^{(1)}_{EP}$ with the TPC.  

Figure~\ref{figura7} shows that the effect of the selection of the angle produces a change of the calculated polarization value. In the future, this fixed distribution of the azimuthal angle will be modeled using results obtained from the relative abundance of the $\Lambda$ and $\bar{\Lambda}$ coming from the high-density core region and a less dense corona, as well as accounting for the intrinsic polarization. We plan to compare with experimental data as well as with data produced with other generators.

\label{sec: Summary}
\section{Summary and outlook}

In this work we have presented a general overview of $\Lambda$ and $\bar{\Lambda}$ reconstruction using the MPD, aimed at measuring the hyperon global polarization for NICA energies and to test the influence of the relative abundances of these particles coming from the core and corona regions of a peripheral heavy-ion collision. The analysis is motivated by a model that has recently been shown to explain the experimental $\Lambda$  and $\bar{\Lambda}$ global polarization differences found at low energies. 

In a future analysis we plan to get the polarization with the measured event plane angle $\Psi^{(1)}_{EP}$ and to improve the selection of $\Lambda$ and $\bar{\Lambda}$ considering the Particle Identification (PID) for the decay product tracks as well as modifying the topological cuts to increase the data set significance. We plan to model the azimuthal angular distributions of the decay baryons to simulate particles coming from the dense and less dense regions of the peripheral heavy-ion collision. We will also compare the results with those obtained using other generators such as DCM-SMM (Statistical Multi Fragmentation Model) and DCM-QGSM (Dubna Cascade Quark Gluon String Model) ~\cite{Baznat:2019iom}.

\section*{Acknowledgments}

I.M. thanks the ICN-UNAM faculty and staff for the support and kind hospitality provided during the development of part of this work. Support for this work has been received in part by UNAM-DGAPA-PAPIIT grant number IG100219 and by Consejo Nacional de Ciencia y Tecnolog\'{\i}a grant numbers A1-S-7655 and  A1-S-16215. I.M. acknowledges support from a postdoctoral fellowship granted by Consejo Nacional de Ciencia y Tecnolog\'{\i}a.

\bibliographystyle{pepan}
\bibliography{pepan_biblio}

\end{document}